\documentclass[preprint2,twocolumn,times,tighten]{aastex631}
\usepackage{subfigure}
\usepackage{newtxtext,newtxmath}
\usepackage{amsmath}
\usepackage{hyperref}
\usepackage{amssymb}
\usepackage{natbib}
\usepackage{morefloats}
\usepackage{multirow}
\usepackage{array}
\usepackage{verbatim}
\usepackage{color}
\let\tablenum\relax
\usepackage{siunitx}

\usepackage{graphicx}	
\usepackage{amsmath}	

\begin{document}

\title{Machine Learning Approach for Estimating Magnetic Field Strength in Galaxy Clusters from Synchrotron Emission}

\author{Jiyao Zhang}
\affiliation{Department of Mathematics, University of Pennsylvania, Philadelphia, PA 19104, USA}

\author[0000-0002-8455-0805]{Yue Hu*}
\affiliation{Institute for Advanced Study, 1 Einstein Drive, Princeton, NJ 08540, USA}

\author{Alex Lazarian}
\affiliation{Department of Astronomy, University of Wisconsin-Madison, Madison, WI 53706, USA}

\email{jiyaoz@sas.upenn.edu; yuehu@ias.edu; alazarian@facstaff.wisc.edu; *NASA Hubble Fellow}



\begin{abstract}
Magnetic fields play a crucial role in various astrophysical processes within the intracluster medium, including heat conduction, cosmic ray acceleration, and the generation of synchrotron radiation. However, measuring magnetic field strength is typically challenging due to the limited availability of Faraday Rotation Measure sources. To address the challenge, we propose a novel method that employs Convolutional Neural Networks (CNNs) alongside synchrotron emission observations to estimate magnetic field strengths in galaxy clusters. Our CNN model is trained on either magnetohydrodynamic (MHD) turbulence simulations or MHD galaxy cluster simulations, which incorporate complex dynamics such as cluster mergers and sloshing motions. The results demonstrate that CNNs can effectively estimate magnetic field strengths with mean squared error of approximately \SI{0.135}{\micro G}$^2$, \SI{0.044}{\micro G}$^2$, and \SI{0.02}{\micro G}$^2$ for $\beta = 100$, 200, and 500 conditions, respectively. Additionally, we have confirmed that our CNN model remains robust against noise and variations in viewing angles with sufficient training, ensuring reliable performance under a wide range of observational conditions. We compare the CNN approach with the traditional magnetic field strength estimates method that assumes equipartition between cosmic ray electron energy and magnetic field energy. Different from the equipartition method, this CNN approach relies on the morphological feature of synchrotron images, offering a new perspective for complementing traditional estimates and enhancing our understanding of cosmic ray acceleration mechanisms.
\end{abstract}



\section{Introduction} \label{sec:intro}
Magnetic fields play a pivotal role across a vast range of astrophysical scales, influencing phenomena from micro-scale cosmic ray transport and acceleration \citep{1966ApJ...146..480J,1978MNRAS.182..443B,2002PhRvL..89B1102Y,2004MNRAS.353..550B,2012SSRv..166...71B,2014ApJ...783...91C,2014ApJ...785....1B,2022ApJ...925...48X,2022ApJ...934..136X,2022MNRAS.512.2111H}, small-scale star formation \citep{1965QJRAS...6..265M, MK04,MO07,2012ApJ...757..154L,2012ApJ...761..156F,2019ApJ...879..129B,HLS21,2022A&A...658A..90A,2023MNRAS.524.4431H}, to large-scale galaxy cluster evolution \citep{2002ARA&A..40..319C,2004IJMPD..13.1549G,2014IJMPD..2330007B,2021MNRAS.502.2518S,2024NatCo..15.1006H}. However, despite its significance, studying magnetic fields, particularly in the Intracluster Medium (ICM), remains notoriously challenging.

Synchrotron emission and polarization serve as key diagnostic tools for estimating the equipartition magnetic field strength and for determining the magnetic field's orientation within the plane-of-the-sky (POS), respectively \citep{1986rpa..book.....R, condon1992radio, 1994ApJ...421..225C, 2009A&A...494...21A, 2016A&A...594A..10P, 2019MNRAS.486.4813Z,2022ApJ...941...92H}. These studies have significantly advanced our understanding, indicating that the typical magnetic field strength in the ICM is at the microgauss (\SI{}{\micro G}) level, with magnetic field correlations extending from a few kpc to hundreds of kpc \citep{2004IJMPD..13.1549G,2005AN....326..414B,2014IJMPD..2330007B,2021MNRAS.502.2518S}. However, the assumption of equipartition between cosmic ray electrons and magnetic fields is poorly justified \citep{2005AN....326..414B}, and the utility of polarized synchrotron emission is often limited by depolarization effects. These include Faraday depolarization, 
i.e., internal depolarization or internal Faraday dispersion, which results from the co-spatial presence of thermal electrons and turbulent magnetic fields along the line-of-sight (LOS) of the emitting regions \citep{brentjens2005faraday}, and beam depolarization, which arises from a turbulent magnetic field distribution within the beam or a uniform foreground medium if the source is emitting at a range of polarization angles. Such depolarization challenges the use of this method in extensive cluster regions, such as radio halos \citep{2004IJMPD..13.1549G,2021MNRAS.502.2518S,2024NatCo..15.1006H}. On the other hand, the Faraday Rotation Measure (RM), derived from the rotation of a polarized source's angle with wavelength, provides insights into the LOS magnetic field, weighted by the thermal electron density \citep{2004A&A...424..429M, melrose2010faraday, brentjens2005faraday}. 
RM based methods have made promising progress on detecting the magnetic field of galaxy clusters over the last two decades \citep{2004A&A...424..429M, 2006A&A...460..425G, 2008A&A...483..699G, 2010A&A...513A..30B, 2010A&A...522A.105G, 2010A&A...514A..71V, 2012A&A...540A..38V, 2016A&A...596A..22B, 2017A&A...603A.122G, 2019MNRAS.487.4768S, 2021MNRAS.502.2518S, 2021PASA...38...20A, 2022A&A...665A..71O, 2025A&A...694A.125L}. Nevertheless, this technique is often limited by the scarcity of radiation sources with well-defined properties, complicating detailed studies \citep{andrecut2011sparse, 2020ApJ...888..101J}.

Most astrophysical fluids are both magnetized and turbulent. Magnetohydrodynamic (MHD) turbulence is anisotropic \citep{GS95,LV99} and this induces the statistical anisotropy of the synchrotron emission \citep{2012ApJ...747....5L}. The elongation direction can be used to probe the POS magnetic field orientation. \cite{2024arXiv240407806H,2024arXiv241009294H} further noticed that this anisotropy not only reflects the POS magnetic field but is also affected by the projection effect along the LOS, and the medium's level of magnetization, defined as $M_A^{-1}$. Here $M_A=\delta v_{\rm inj}\sqrt{4\pi\rho}/B$ is the Alfv\'en Mach number, with $\delta v_{\rm inj}$ being turbulence injection velocity, $\rho$ mass density, and $B$ magnetic field strength. \cite{2024arXiv240407806H} introduced the use of Convolutional Neural Networks (CNNs; \citealt{lecun1998gradient}) to extract the observed anisotropy in synchrotron emissions and thereby facilitate the measurement of 3D magnetic fields. 

In this paper, we aim to extend the CNN approach to estimate magnetic field strength using synchrotron emission. \cite{2024arXiv240407806H} primarily retrieved the magnetization of the medium rather than its magnetic field strength. Deriving the magnetic field strength from $M_A$ requires the information of turbulence injection velocity, which is not available in diffuse ICM. \cite{MsCNN} extends the CNN approach to estimate the kinetic energy of turbulence, characterized by the sonic Mach number $M_s=\delta v_{\rm inj}/c_s$, where $c_s$ is the sound speed. The product $M_A^{-1} M_s$ thus gives the (square-root) ratio of magnetic field energy and thermal kinetic energy. It suggests that the two Mach numbers contain information on magnetic field strength \citep{2022ApJ...935...77L,2024ApJ...974..237L}. Therefore, we aim to develop a CNN model to extract the strength information directly. Importantly, this approach does not assume equipartition between cosmic ray electrons and magnetic fields, allowing for a comparison with traditional equipartition estimates and providing insights into the cosmic ray acceleration mechanism.

Our research extends the scope of previous research in the field. Notably, the training of CNNs in \cite{2024arXiv240407806H} was based on simulations of MHD turbulence and did not include complexities like cluster mergers, sloshing motions, or cooling and heating dynamics found in galaxy clusters, which might affect the validity of their model under realistic cluster conditions. In this work, we trained and tested the CNN model with numerical MHD galaxy cluster simulations (Fig.~\ref{intensity_b}) from the Galaxy Cluster Merger Catalog \citep{2011ApJ...743...16Z,2018ApJS..234....4Z}. We evaluated the accuracy of the estimation and further quantitatively tested the robustness of the model with rotation angle and noise. We also compared the result with the equipartition magnetic field strength. These cluster simulations feature three different initial plasma compressibilities ($\beta = 100, 200, 500$) and include a dark matter subcluster passing near the center of a main cool-core cluster, mimicking the Perseus cluster.

This paper is structured as follows: in \S~\ref{sec:theory} we discuss the theoretical consideration that the morphology of synchrotron emission observation contains information on magnetic field strength. In \S~\ref{sec:data}, we detail the MHD turbulence and galaxy cluster simulations, as well as the mock synchrotron observation, utilized in our study. \S~\ref{sec:method} briefly revisits the architecture of our CNNs and our training strategies. In \S~\ref{sec:results}, we present the results derived from our numerical testing, offering insights into the efficacy and accuracy of the CNN model. \S~\ref{sec:dis} delves into discussions surrounding the uncertainties and prospects of employing CNNs for observational analysis. We conclude with a summary of our findings in \S~\ref{sec:con}.



\section{Theoretical consideration}
\label{sec:theory}
To illustrate the physics behind the CNN approach for estimating the magnetic field strength, we start with a clean system considering only the MHD turbulence.

\subsection{Magnetic field strength is imprinted in $M_A$ and $M_s$}
Alfv\'en Mach number $M_A$ and sonic Mach number $M_s$ are two important parameters in characterizing the properties of MHD turbulence. The two Mach numbers are defined as:
\begin{align}
\label{eq.Ma}
M_A &= \frac{\delta v_{\rm inj}}{v_A} = \frac{\delta v_{\rm inj}}{B}\sqrt{4\pi\rho},\\
M_s &= \frac{\delta v_{\rm inj}}{c_s},
\end{align}
where $v_A$ is the Alfv\'en speed. The squared $M_A$ reflects the relative importance of turbulent kinetic energy and magnetic field energy, while the squared $M_s$ characterizes the ratio of turbulent kinetic energy and thermal kinetic energy.

Combining two Mach numbers, the magnetic field strength $B$ can be expressed as:
\begin{equation}
\label{eq.B}
    B=c_s\sqrt{4\pi\rho}M_sM_A^{-1},
\end{equation}
which suggests that if $M_A$ and $M_s$ are known, one could determine magnetic field strength from observations. This approach, denoted as MM2, was originally suggested and elaborated by \cite{2022ApJ...935...77L,2024ApJ...974..237L}. 

\begin{figure*}[htbp!]
    \centering
    \includegraphics[width=1.0\linewidth]{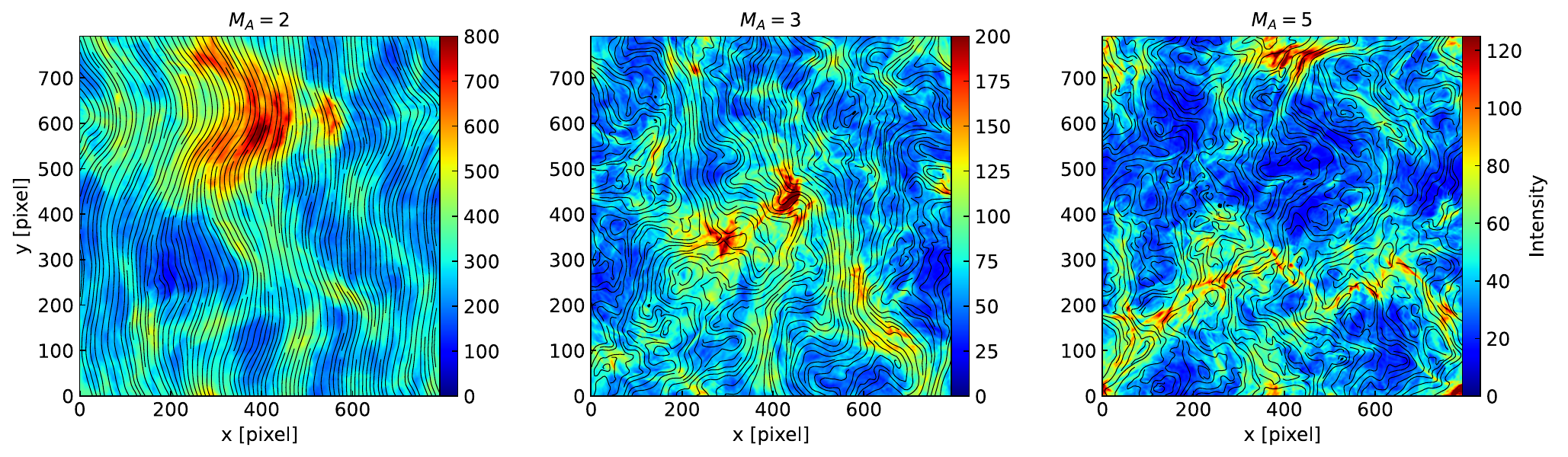}
    \caption{Maps of synchrotron emission intensity generated from MHD turbulence simulations. Three different initial $M_A=2$ (left), $M_A=3$ (middle), and $M_A=5$ (right) are presented. The black streamlines represent the density-weighted POS magnetic field orientation. }
    \label{fig.tur}
\end{figure*}

\subsection{Estimating $M_A$ with CNN}
Estimating $M_A$ with CNN and synchrotron emission is proposed in \cite{2024arXiv240407806H}. For a turbulent eddy at scale $l$, we can decompose its scale into the parallel $l_\parallel$ and perpendicular $l_\bot$ components with respect to the local magnetic field. Turbulent energy cascades predominantly in the perpendicular direction due to fast turbulent reconnection, which minimizes resistance in that direction. Eventually, this cascade achieves a state of "critical balance" with the Alfv\'en wave, which propagates along the parallel direction. This balance can be expressed as:
\begin{equation}
\label{eq.cb}
    \delta v_{l_\bot}l_\bot^{-1}\sim v_Al_\parallel^{-1},
\end{equation}
where $\delta v_{l_\bot}$ denotes the velocity fluctuations at the perpendicular scale $l_\bot$. Additionally, in the regime of strong turbulence, the cascade follows a Kolmogorov-type scaling relation \citep{xu2019study}: 
\begin{equation}
\label{eq.kl}
     \delta v_{l_\bot} = (\frac{l_\bot}{L_{\rm inj}})^{1/3}v_{\rm inj}M_A^{1/3},
\end{equation}
where $L_{\rm inj}$ is the injection scale $L_{\rm inj}$. Combing Eqs.~\ref{eq.cb} and \ref{eq.kl}, one can get the scale-dependent anisotropy scaling \citep{LV99}\footnote{This scaling relation is valid for sub-Alfv\'enic turbulence with $M_A\le1$. For super-Alfv\'enic scenarios where $M_A \gg 1$, turbulence cascades energy from larger injection scales down to smaller scales and progressively diminishes turbulent velocity. Assuming Kolmogorov turbulence, the magnetic field's energy approaches that of turbulence (i.e., the $M_A$ is unity) at the transition scale $l_A=L_{\rm inj}/M_A^3$, below which the magnetic field's role becomes important and the anisotropy can be observed \citep{2006ApJ...645L..25L}. $l_A$ is estimated to be 1 - 60 kpc in ICM \citep{2024NatCo..15.1006H}.}:
\begin{align}
 l_\parallel= L_{\rm inj}(\frac{l_\bot}{L_{\rm inj}})^{\frac{2}{3}} M_A^{-4/3},
\label{eq.lv99}
\end{align}
which suggests that the ratio of $l_\parallel/l_\bot$ depends on $M_A$. A higher magnetization leads to a larger $l_\parallel/l_\bot$ ratio, while weaker magnetization decreases this ratio.

Eq.~\ref{eq.lv99} provides the scaling relation for velocity fluctuations. The relationships for magnetic field fluctuations $\delta B_l$ can also be derived using the linearized continuity and induction equations, considering the components as a sum of their mean and fluctuating parts (see \citealt{2024arXiv240407806H} 
 for details): 
\begin{align}
\delta B_l&= \delta v_l\frac{B}{v_A}\mathcal{F}^{-1}(|\hat{\pmb{B}}\times\hat{\pmb{\xi}}|), 
\label{eq.rhoB}
\end{align}
where $\hat{\pmb{k}}$ and $\hat{\pmb{\xi}}$ represent the unit wavevector and displacement vector, respectively. $\mathcal{F}^{-1}$ denotes the inverse Fourier transform. Since fluctuations in synchrotron emission intensity primarily arise from magnetic field fluctuations (see Eq.~\ref{eq.ip}), their statistical properties are governed by MHD turbulence. Consequently, the aspect ratio of an observed synchrotron structure correlates with $M_A$ (see Fig.~\ref{fig.tur}).

However, the observed synchrotron intensity is projected onto the POS. The projection alters the $l_\parallel/l_\bot$ ratio. This introduces a degeneracy, as both LOS projection and changes in $M_A$ can affect the observed anisotropy. To resolve this degeneracy, additional information on the structure's morphological curvature is required. Strong magnetic field fluctuations, i.e., large $M_A$, result in significantly curved magnetic field lines (see Fig.~\ref{fig.tur}). These fluctuations are naturally mirrored in synchrotron observations and their morphological curvature. This provides key information for estimating $M_A$. \cite{2024arXiv240407806H} proposed utilizing CNNs to extract the information and ultimately estimate $M_A$ from synchrotron emission data.

\begin{table*}[htbp!]
\centering
\resizebox{0.75\linewidth}{!}{
\centering
\begin{tabular}{|c|c|c|c|c|}
\hline
Simulation Names & Initial $\beta$ & Resolution & Minimum B-strength & Maximum B-strength \\ \hline
beta100\_310     & 100                          & 2  kpc / pixel             & \SI{0.010}{\micro G}                       & \SI{5.57}{\micro G}                                  \\ \hline
beta200\_310     & 200                          & 2  kpc / pixel             & \SI{0.004}{\micro G}                        & \SI{4.75}{\micro G}                                  \\ \hline
beta500\_310     & 500                          & 2  kpc / pixel             & \SI{0.002}{\micro G}                         & 
\SI{3.13}{\micro G}                                 \\ \hline
\end{tabular}}
\caption{Overview of galaxy cluster simulations used in this work. We use the snapshots of the clusters at 3.1 Gyr. Three different initial plasma $\beta$ are included.}
\label{tab:sim}
\end{table*}

\subsection{Estimating $M_s$ with CNN}
In addition to estimating $M_A$, Schmaltz et al. (2024, in prep.) extend the CNN to measure $M_s$. The relationship between $M_s$, density fluctuations, and magnetic field fluctuations is well-established. In media with high $M_s$, shocks play a more dominant role, and the small-scale, high-amplitude fluctuations produced by these shocks are imprinted in synchrotron intensity maps. This indicates that variations in $M_s$ are distinctly reflected in the intensity maps, providing a physical foundation for extracting $M_s$ information from synchrotron observations using CNNs. For further details, we refer readers to \cite{MsCNN} .

To summarize, the observed synchrotron emission map contains detailed information on $M_A$ and $M_s$. Especially, the morphological features, including the aspect ratio, curvature, and small-scale shock structures, in the synchrotron map are sensitive to either $M_A$ or $M_s$. The combination of $M_A$ and $M_s$, on the other hand, correlates with the magnetic field strength (see Eq.~\ref{eq.B}). In Appendix~\ref{appendix}, we provide a test of the CNN's performance in estimating the magnetic field strength using MHD turbulence simulations. In a complicated system of galaxy clusters, fluctuations in synchrotron intensity are not only induced by MHD turbulence. However, the changes in magnetic field strength, thus, are also expected to be imprinted in the synchrotron map, as shown in \S~\ref{sec:results}.

\section{Numerical Simulations} \label{sec:data}

\begin{figure*}[htbp!]
    \centering
    \includegraphics[width=1.0\linewidth]{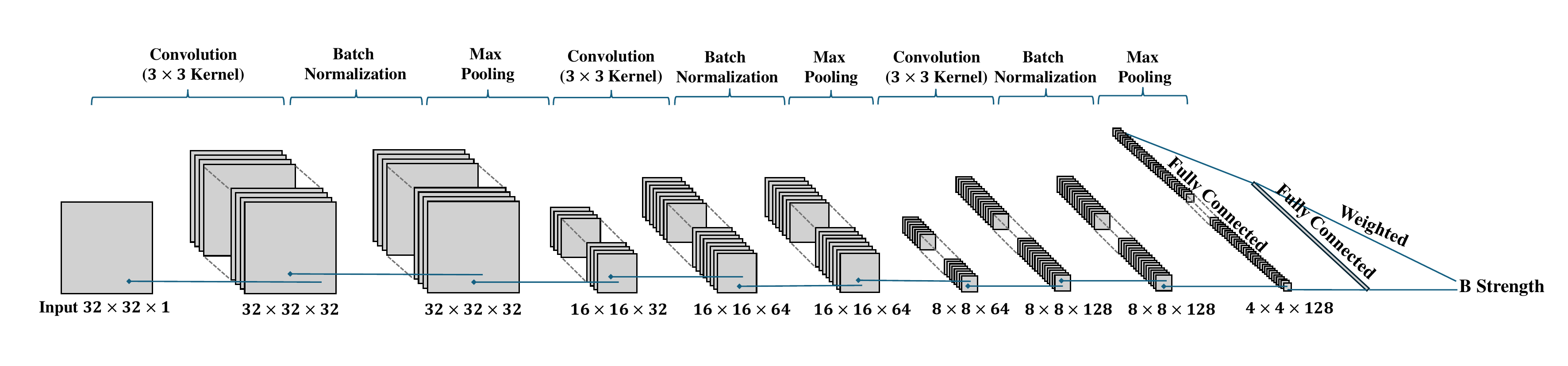}
    \caption{Architecture of the CNN-model. The input image is a $32 \times 32$ pixel map cropped from the synchrotron intensity map. The network outputs a scalar - the prediction of magnetic field strength. Modified from \protect\citet{2024MNRAS.52711240H}.}
    \label{cnn_structure}
\end{figure*}
\subsection{MHD simulations of galaxy cluster}
The numerical simulations used in this research come from \cite{2011ApJ...743...16Z}.
They simulated idealized binary galaxy cluster mergers by solving ideal MHD equations and setting up a dark matter subcluster passing near the center, mimicking typical conditions of cool-core clusters and producing sloshing and turbulent gas motions in cluster cores. The simulations were generated from the parallel hydrodynamics/N-body astrophysical simulation code FLASH 3 \citep{fryxell2000flash, dubey2009extensible}. Using a directionally unsplit staggered mesh (USM) algorithm which guarantees that the evolved magnetic field satisfies the divergence-free condition \citep{evans1988simulation}. 

Three distinct initial conditions were tested, characterized by different plasma compressibilities: $\beta=100,200,500$, where $\beta=p_{\rm th}/p_B$. Here, $p_{\rm th}$ and $p_B$ represent the gas thermal pressure and magnetic field pressure, respectively. These conditions encompass a range of initial magnetic field strengths, as detailed in Tab.~\ref{tab:sim}. The simulated galaxy cluster spans a length scale of 1 Mpc and is discretized into a $512^3$ grid, achieving a resolution of approximately 2 kpc per pixel. For additional details regarding the setup of these simulations,  we refer readers to \cite{2011ApJ...743...16Z}.
  

\subsection{Synthetic synchtron emission}
To generate synthetic synchrotron observations from our simulations, we use the density field
$\rho(\pmb{x})$ and the magnetic field $\pmb{B}(\pmb{x})$. The calculations for synchrotron intensity 
$I(\pmb{X})$ is based on \citep{1986rpa..book.....R,1970ranp.book.....P,2016ApJ...831...77L}:
\begin{equation}
I(\pmb{X}) \propto \int dz~n_{e,r} (B_x^2+B_y^2)^{\frac{\gamma-3}{4}} (B_x^2+B_y^2),
\label{eq.ip}
\end{equation}
where $\pmb{X} = (x,y)$ and $\pmb{x} = (x,y,z)$ represent spatial coordinates. $n_{e,r}(\pmb{x})=\rho(\pmb{x})$ is the relativistic electron number density, and $\gamma$ denotes the spectral index of the electron energy distribution $E$:
\begin{equation}
\label{eq.NE}
N(E) dE = N_0 E^{-\gamma}dE,
\end{equation}
with $N(E)$ representing the electron number density per unit energy interval $dE$. The pre-factor $N_0$ is derived by integrating Eq.~\ref{eq.NE} to obtain the total electron number density. Given the relative insensitivity of synchrotron emission to variations in the electron energy distribution's spectral index, as noted by \cite{2012ApJ...747....5L, 2019ApJ...886...63Z}, we assume a homogeneous and isotropic electron energy distribution with a spectral index $\gamma=3$. Other constant factors at a given $\gamma$ are not explicitly detailed in Eq.~\ref{eq.ip}, as they do not alter the characteristics of the synchrotron intensity fluctuations.

\section{Methodology} 
\label{sec:method}

\subsection{CNN Model}
The CNN model employed in this study follows the architecture outlined by \citet{2024MNRAS.52711240H}, specifically designed to interpret the 3D magnetic fields from synchrotron emission maps. The detailed architecture of this model is illustrated in Fig.~\ref{cnn_structure}, showcasing a multi-layered approach that includes convolutional layers, batch normalization layers, and max pooling layers.

\textbf{Convolutional Layers:} These layers utilize various convolution kernels to separate and extract local features from the input images. Each convolutional layer aims to identify distinct patterns within the data, which are crucial for understanding the underlying magnetic field structures.

\textbf{Batch Normalization Layers:} Following each convolutional layer, a batch normalization layer is implemented. This layer normalizes the activations of the previous layer at each batch, i.e., it applies a transformation that maintains the mean output close to 0 and the output standard deviation close to 1. This normalization helps to expedite convergence and enhance learning stability during network training via backpropagation \citep{ioffe2015batch}.

\textbf{Max Pooling Layers:} A common variant in the pooling layers is the Max Pooling Layer, which reduces the spatial size of the input images, making the detection of features invariant to scale and orientation changes more efficient \citep{scherer2010evaluation}. This layer operates by selecting the maximum value from a specified window of neurons and outputs this value, effectively reducing the dimensionality of the input feature map while retaining the most critical feature information.

After several cycles of convolution, normalization, and pooling, the dimensions of the input images are significantly reduced, yet enriched with important local features. These processed images are then flattened into a 1D vector containing the extracted features. This vector feeds into the fully connected layers, which perform the final analysis to predict the strength of the magnetic field based on the learned features. A detailed discussion of each layer’s function is given in \cite{2024MNRAS.52711240H}. 

\subsection{Training Strategy}
The CNN architecture described has demonstrated its effectiveness in tracing 3D magnetic fields using synchrotron observations. The model’s trainable parameters are optimized using a typical conventional neural network training approach. The optimization process is guided by the Mean-Squared Error (MSE) of the predicted magnetic field strength distribution against the actual field strength, which serves as the loss function for backpropagation. This methodology is based on the foundational principles laid out by \citep{rumelhart1986learning}, which emphasize the importance of efficient error reduction through iterative learning.

The input image is normalized synchrotron emission maps. The normalization ensures that only the morphological features of the synchrotron structure, rather than its intensity, are captured for training. To further enhance the model’s generalization capabilities and ensure robust performance on unseen data, we have implemented a data augmentation strategy during the training phase \citep{Takahashi_2020}. This strategy involves subjecting the input synchrotron images to random cropping, reducing them to patches of 32$\times$32 pixels. Each of these patches is then rotated by a random angle before being fed into the CNN. This process of cropping and rotating not only introduces variability but also injects randomness into the training dataset \citep{van2001art, Larochelle2007}. The network outputs absolute magnetic field strengths.

\begin{figure*}[htbp!]
    \centering
    \includegraphics[width=\linewidth]{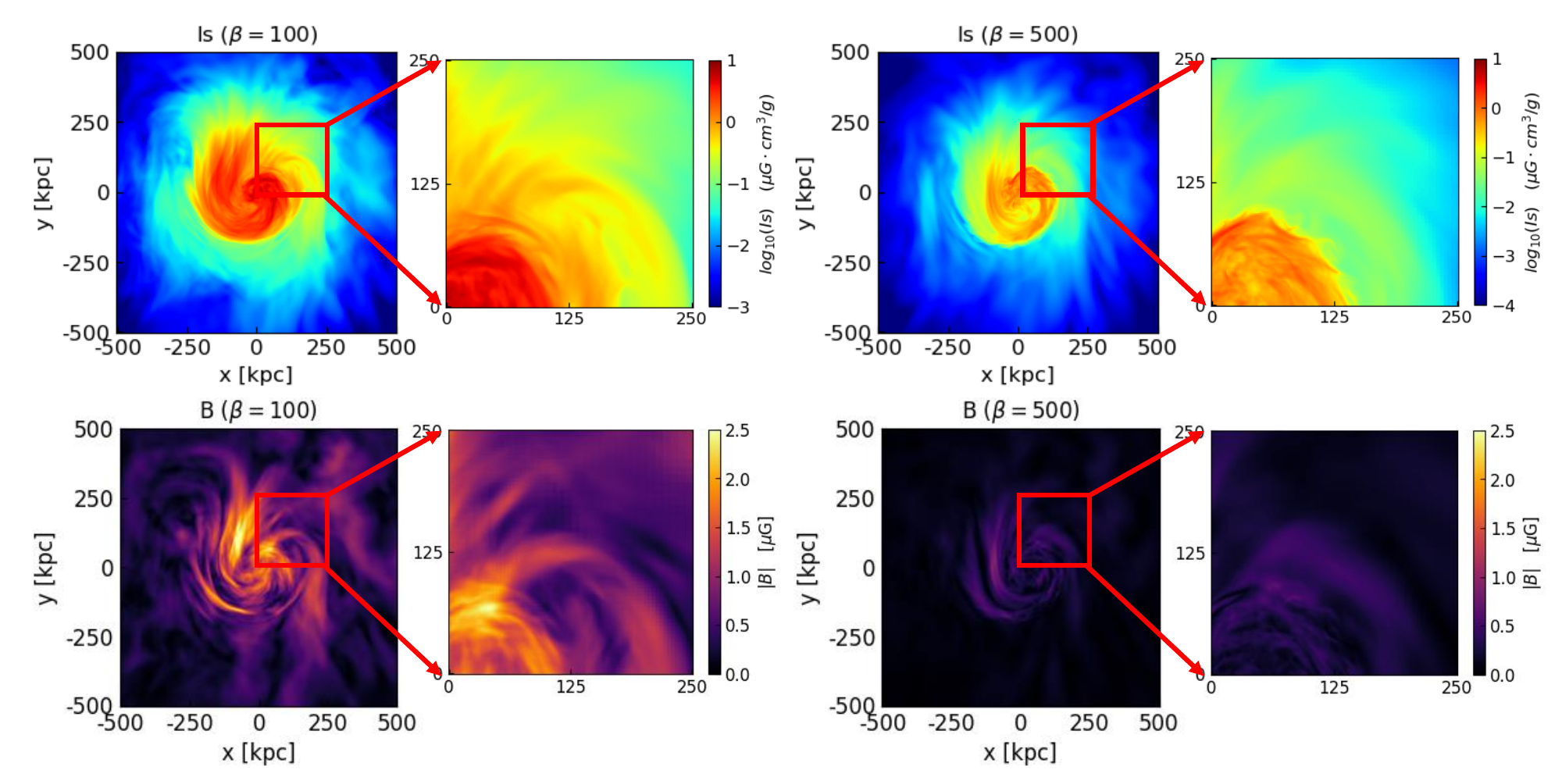}
    \caption{Maps of synchrotron emission intensity (top) and the corresponding project mass-weighted magnetic field strength (bottom). Two different initial $\beta = 100$ (left) and $\beta = 500$ (right) are presented. The zoom-in image on the right of each panel shows the structural differences in synchrotron intensity maps due to different initial $\beta$.}
    \label{intensity_b}
\end{figure*}

\begin{figure*}{}
    \centering
    \includegraphics[width=0.8\linewidth]{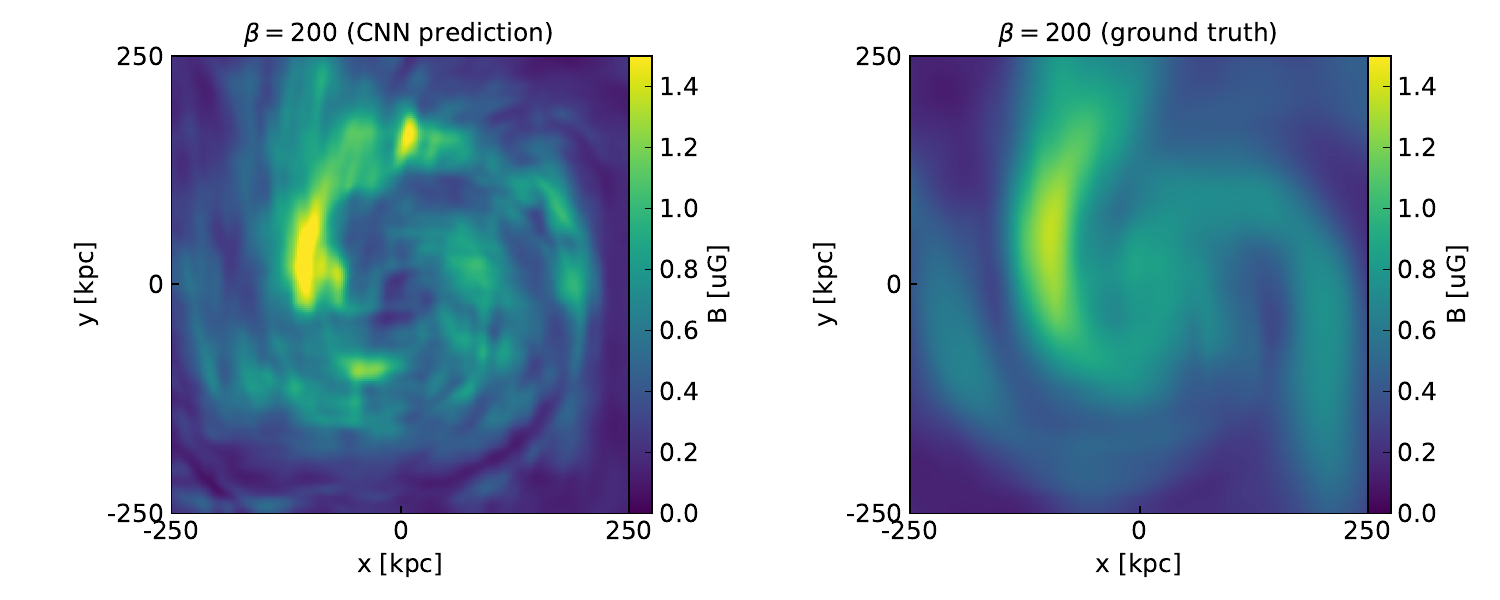}
    \caption{A comparison of the magnetic field strength predicted by CNN (left panel) and the actual mass-weighted magnetic field strength (right panel) in the $\beta=200$ simulation. }
    \label{comparison}
\end{figure*}

\begin{figure*}
    \centering
    \includegraphics[width=1.0\linewidth]{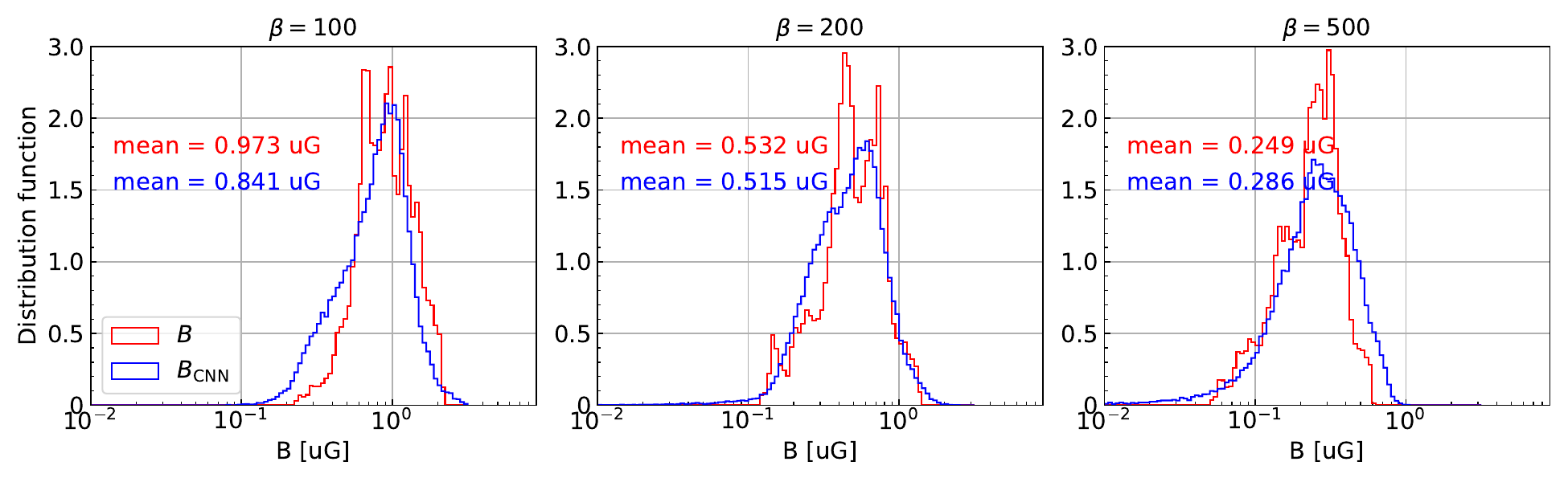}
    \caption{\label{predictions} Histograms of the CNN-predicted magnetic field strength (red) and the corresponding actual density-weighted $B$  (blue). The three panels represent three physical conditions of $\beta=100$ (left), $\beta=200$ (middle), and $\beta=300$ (right).}
\end{figure*}

\section{Results}
\label{sec:results}
\subsection{Synchrotron emission's morphological correlation with magnetic field strength}
The morphology observed in synchrotron emission maps is intricately influenced by both the distribution of relativistic electron number density and the magnetic field projected onto the POS (see Eq.~\ref{eq.ip}). A stronger magnetic field enhances magnetic pressure, compressing the gas or particles and leading to synchrotron structures that appear less clumped compared to those formed in regions with weaker magnetic fields. Moreover, magnetic-field-dependent instabilities, such as the magnetorotational instability (MRI; \citealt{2015JPlPh..81e4908N}) and the magneto-thermal instability (MTI; \citealt{2022MNRAS.513.4625P}), generate turbulence in galaxy clusters. This turbulence further affects the distribution of gas density and magnetic field fluctuations, which in turn shape the observed synchrotron structures, providing insights into the POS magnetic field strength \citep{2024arXiv240407806H,2024arXiv241009294H}.

An example can be seen in Fig.~\ref{intensity_b}, which compares synchrotron emission maps across different magnetic field strengths. For a strong magnetic field case ($\beta=100$), the synchrotron map displays many filamentary and spur-like small-scale structures, particularly prominent in the central regions of the cluster. Conversely, such filamentary and spur-like structures are markedly less pronounced in a scenario with a weaker magnetic field ($\beta=500$). This morphological difference underpins the rationale for employing a CNN model to predict magnetic field strength, as machine learning models are particularly adept at capturing and extracting these spatial features.


\subsection{CNN training and test using galaxy cluster simulations}
For training the CNN model, we employed data augmentation techniques including random cropping and rotation of the input images. Specifically, sub-fields of 32$\times$32 pixels were randomly extracted from synchrotron emission maps corresponding to three different magnetic field strengths (with $\beta = 100$, 200, and 500). Approximately 0.5 million sub-fields were used in each training iteration, and the target variable was the density-weighted total magnetic field strength. The model was trained over 30 iterations, continuing until the mean squared error loss function saturated, indicating effective convergence of the network parameters. Following training, the original synchrotron emission maps were used to evaluate the accuracy of the CNN. Notably, the randomly rotated images have matrix representations that differ from those of the original images; from the CNN’s perspective, these represent distinct data points, thereby enhancing the model’s robustness to orientation variations.

Fig.~\ref{comparison} presents a comparison between the CNN-predicted magnetic field strength and the actual mass-weighted magnetic field strength in the $\beta=200$ simulation\footnote{Results from the CNN trained with turbulence simulations are provided in Appendix~\ref{appendix}.}. In this figure, the actual magnetic field map has been smoothed to match the resolution of the CNN prediction. The smoothing removes very strong magnetic field spots seen in Fig.~\ref{intensity_b}. Here, we focus on the central 500~kpc regions where the dynamics of the galaxy cluster are most prominent. Since there is no external turbulence driving or additional dynamical processes—other than galaxy mergers—the outskirts exhibit very weak magnetic fields and introduce noise-like structures. Despite some apparent differences between the CNN prediction and the ground truth, a global similarity exists.

Furthermore, Fig.~\ref{predictions} displays histograms comparing the CNN predictions, denoted as $B_{\rm CNN}$, with the actual magnetic field strengths $B$ derived from the simulations for $\beta = 100$, 200, and 500. These values span the typical magnetic field strength range observed in galaxy clusters \citep{2002ARA&A..40..319C,2004IJMPD..13.1549G}. The histograms of $B_{\rm CNN}$ and $B$ are similar for $\beta = 500$ and 200; however, as $\beta$ increases, $B_{\rm CNN}$ extends up to \SI{0.1}{\micro G} , deviating somewhat from the $B$ distribution. In addition, the Mean Squared Error (MSE) values computed for $B_{\rm CNN}$ versus $B$ are given in Tab.~\ref{tab:mse}. The minimum MSE value of 0.02 corresponds to the $\beta = 500$ case, indicating that despite some deviations, the predictions statistically correlate well with the actual magnetic field strengths. This confirms the model's capability to estimate magnetic field strength based on synchrotron emission maps.

\begin{figure*}[htp!]
    \centering
    \includegraphics[width=1.0\linewidth]{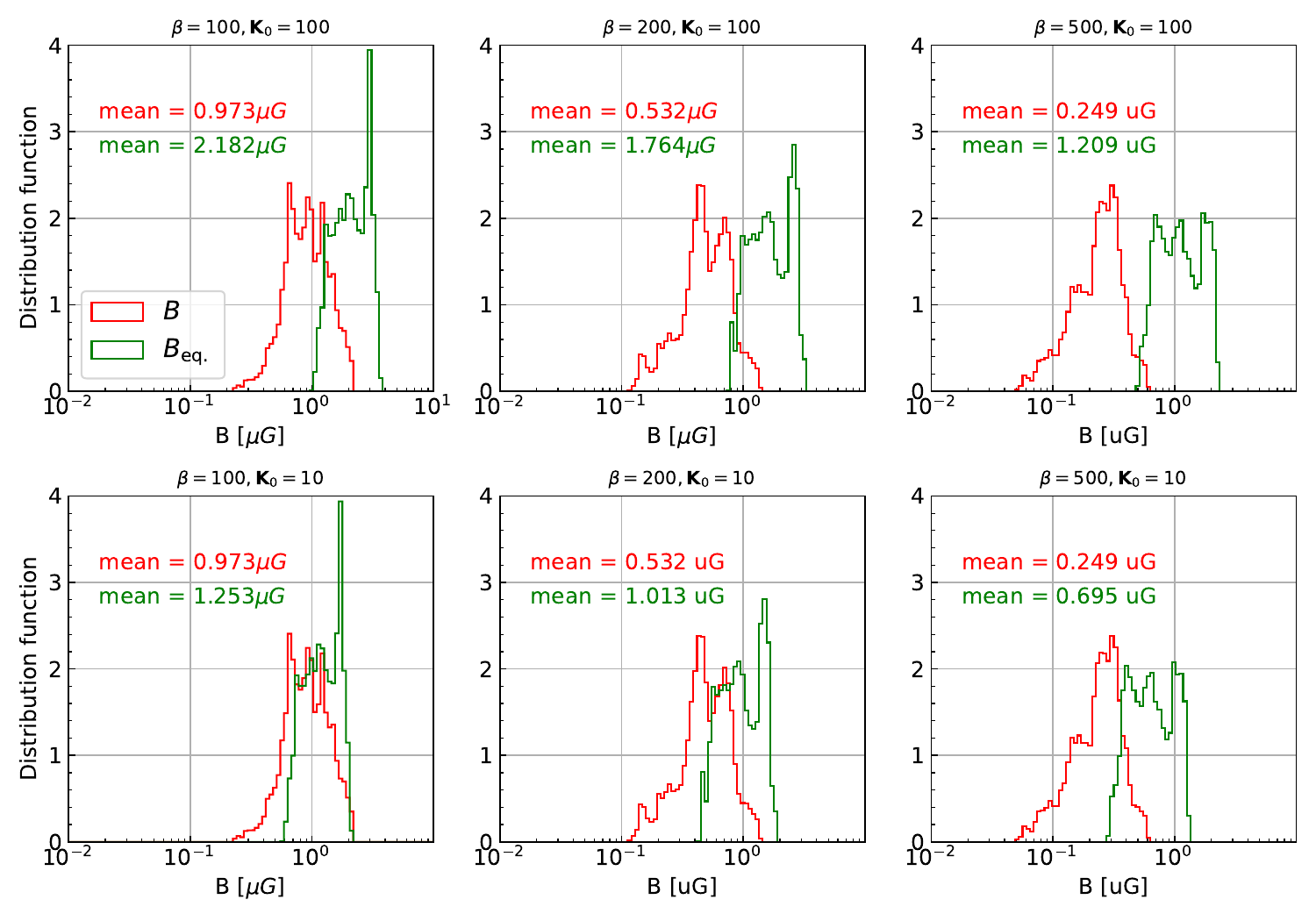}
    \caption{Histograms of the equipartition magnetic field strength $B_{\rm eq}$ and the corresponding actual density-weighted strength $B$. The three panels represent three physical conditions of $\beta=100$ (left), $\beta=200$ (middle), and $\beta=300$ (right).  $\pmb{\rm K}_0$ is the number density ratio of cosmic ray protons and electrons per particle energy interval. 
    }
    \label{fig.Beq}
\end{figure*}

\begin{table*}
	\centering
\begin{tabular}{ | c | c | c | c | }
    \hline
     & MSE for $\beta=100$ [\SI{}{\micro G} $^2$] & MSE for $\beta=200$ [\SI{}{\micro G} $^2$] & MSE for $\beta=500$ [\SI{}{\micro G} $^2$] \\ \hline \hline
    $B_{\rm CNN}$ & 0.135 & 0.044 & 0.020 \\ 
    $B_{\rm eq}$ ($\pmb{\rm K}_0=10)$ & 0.190 & 0.313 & 0.244 \\ 
    $B_{\rm eq}$ ($\pmb{\rm K}_0=100)$ & 1.090 & 1.792 & 1.729 \\ 
    \hline
\end{tabular}
    \caption{\label{tab:mse} Mean Squared Error (MSE) values of the CNN prediction $B_{\rm CNN}$ and equipartition magnetic field strength $B_{\rm eq}$. $\pmb{\rm K}_0$ is the number density ratio of cosmic ray protons and electrons per particle energy interval.}
\end{table*}

\subsection{Comparison with the equipartition magnetic field strength}
To estimate the magnetic field strength using radio observation, an equipartition between magnetic fields and cosmic rays is assumed.  The CNN-based approach, however, relies on the morphological feature of synchrotron images. By comparing traditional methods with the CNN-based approaches, we can better quantify the energy distribution or equipartition in galaxy clusters—a crucial factor in understanding the history and evolution of cosmic ray acceleration.

We make a comparison of the equipartition method and the CNN-based approach to measure magnetic field strength. This method assumes the energy in magnetic fields is $U_B \propto B^2$, the energy in relativistic particles is $U_{\rm part} = U_{\rm el} + U_{\rm pr} \propto B^{-3/2}$. 
The total energy content $U_{\rm tot}$ is minimum when the contributions of magnetic fields and relativistic particles are approximately equal (equipartition condition). 
The corresponding magnetic field is commonly referred to as the equipartition value $B_{\rm eq}$.
Here we adopt the revised equipartition equation given by \cite{2005AN....326..414B} (see their Appendix A for more details), in which $B_{\rm eq}$ is given by the following:
\begin{equation}
\label{update_eq}
B_{\rm eq} = (\frac{4\pi(2\alpha+1)(1+\pmb{\rm K}_0)I_\nu E_p^{1-2\alpha}(\frac{\nu}{2C_1})^{\alpha}}{(2\alpha-1)C_2lC_4})^{1/(\alpha+3)},
\end{equation}
where $\alpha = (\gamma-1)/2$ and $\gamma$ is the injection spectral index of the energy spectrum. $l$ represents the integration pathlength and $E_p = 938.28$~Mev is the pectral break energy for protons. $C_1=6.26\times10^{18} {\rm erg}^{-2}{\rm s}^{-1}{\rm G}^{-1}, C_3=C_3(\gamma+7/3)/(\gamma+1)\Gamma[(3\gamma-1)/12]\times\Gamma[(3\gamma+7)/12)], C_3=1.86\times10^{-23}~{\rm erg}~{\rm G}^{-1}~{\rm sterad}^{-1}$ are constants. 

$\pmb{\rm K}_0=n_{p.0}/n_{e,0}$ refers to the number density ratio of cosmic ray protons and electrons per particle energy interval within the energy range. Following the synchrotron formula \citep{1970ranp.book.....P}, the revised synchrotron intensity observed as the frequency $\nu$, $I_\nu$ is given by:
\begin{equation}
\label{LocalSynI}
I_v = \int dz~C_2N_0E_0^{\gamma}(\frac{\nu}{2C_1})^{(1-\gamma)/2}B^{(1+\gamma)/2}C_4,
\end{equation}
where $C_4=(2/3)^{(\gamma+1)/4}$ accounts for the projection of total magnetic field strength to POS magnetic field strength. Compared with the classical equipartition equation (Eq.~\ref{eq.ip}), the revised method considers a particular energy limit $E_p$, referring to the point where state function for sub-relativistic cosmic ray electrons and photons becomes flattened due to ionization and/or Coulomb loss, resulting in a piece-wise particle density function \citep{pohl1993predictive}.

In our calculations, we assume a frequency of $\nu=1$ GHz and consider values of $\pmb{\rm K}_0 = 10$ and $\pmb{\rm K}_0 = 100$. Fig.~\ref{fig.Beq} presents histograms that compare the revised equipartition magnetic field strength, $B_{\rm eq}$ (Eq.~\ref{update_eq}), with the actual total magnetic field strength $B$. The results show that $B_{\rm eq}$ is highly sensitive to the assumed value of $\pmb{\rm K}_0$, which represents the ratio of cosmic ray proton to electron number densities. A larger $\pmb{\rm K}_0$ implies that less energy is allocated to electrons, necessitating a stronger magnetic field to achieve the same level of synchrotron emission. Consequently, $B_{\rm eq}$ tends to overestimate the magnetic field strength when $\pmb{\rm K}_0$ is large, while a smaller $\pmb{\rm K}_0$ can lead to an underestimation.

In comparison to the CNN prediction (as quantified in Tab.~\ref{tab:mse}), $\pmb{\rm K}_0$ is a critical yet challenging parameter to estimate in observational data accurately. The CNN approach, on the other hand, relies solely on the morphological features in synchrotron emissions, enabling us to estimate the electron-to-proton ratio by comparing the CNN-predicted magnetic field with $B_{\rm eq}$. Nevertheless, a more detailed investigation of the parameter space is needed to make the synergy of the two methods.

\begin{figure*}
    \begin{subfigure}   
    \centering
    \includegraphics[width=0.99\linewidth]{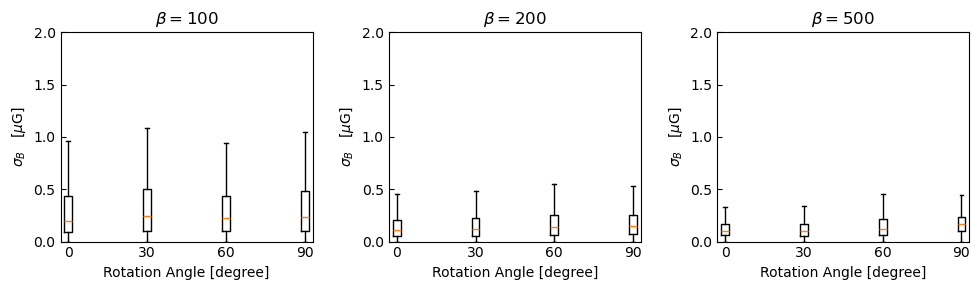}
   \end{subfigure}
    \centering
    \begin{subfigure}   
    \centering
    \includegraphics[width=0.99\linewidth]{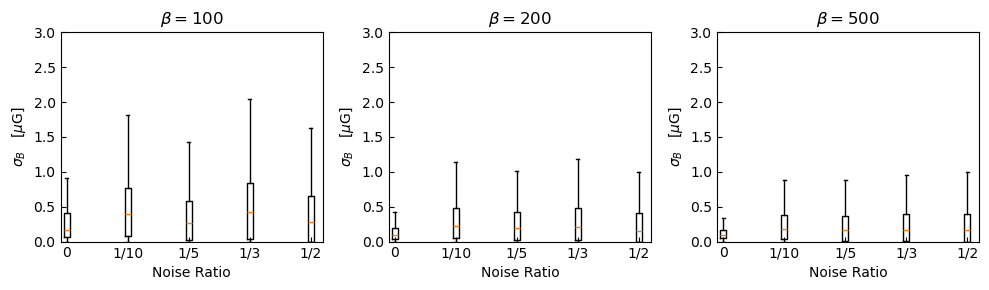}
   \end{subfigure}
    \caption{Top: Boxplot of the errors of CNN model predictions with certain angle rotation in training data, i.e. $\beta = 100$ (left), $\beta = 200$ (middle), and $\beta = 500$ (right), and the corresponding rotation angle with 0, 30, 60, and 90 degrees. Bottom: Boxplot of the errors of CNN model predictions with certain noise ratio in training data, i.e. $\beta = 100$ (left), $\beta = 200$ (middle), and $\beta = 500$ (right), and the corresponding noise ratio with 0, 1/10, 1/5, 1/3, and 1/2.  }
    \label{fig:boxplot}
\end{figure*}



\section{Discussion}
\label{sec:dis}

\subsection{Uncertainty}
\textbf{Uniform spectral index:} The training of the CNN model in this study relies on MHD simulations of turbulence (see Appendix~\ref{appendix}) or galaxy cluster mergers. To generate synthetic synchrotron observations, we assumed a uniform spectral index. In practice, however, the spectral index may vary across synchrotron emission maps, as electrons in different regions may undergo distinct acceleration and cooling processes. The variations in spectral index may introduce additional fluctuations, potentially altering the morphology of synchrotron structures. Despite these complexities, we anticipate that the CNN model can capture complex correlations with magnetic field strength as more detailed synchrotron emission simulations become available for training.

\textbf{Generalization:} While the Perseus cluster provides valuable insights into the magnetic structure of a typical cool-core cluster, it may not represent the full diversity of galaxy clusters. The dynamics of turbulence, gas accretion, and field amplification processes may differ substantially across galaxy clusters, introducing uncertainties when applying the model to more varied or complex scenarios. A more detailed exploration of the parameter space is needed to generalize the CNN-based approach and create synergy with the equipartition method.

\textbf{Observational effect:} Model performance could be significantly degraded with real-world observations due to uncertainties in observational effects. Deconvolution, a critical step in reconstructing detailed magnetic field maps from observational data, amplifies distortion and noise signals from external interference of terrestrial or satellite sources, especially in regions with weak or complex signal patterning. The presence of such interference complicates the deconvolution process, as distinguishing between astrophysical informative signals and Radio Frequency Interference (RFI) becomes challenging. Here we conduct two tests for different view angles and noise.

\subsubsection{Effect of different view angles}
A remarkable challenge in galaxy observation is the angle of observation. Galaxy clusters are not always face-on (see Fig.~\ref{intensity_b}), and the projection effect could change the observed synchrotron emission and the projected magnetic field strength. Therefore, we introduced a certain rotation angle to the training data to evaluate the performance of the CNN model when the cluster is not face-on (i.e., rotation angle $>0$ degrees). We rotate the simulation cubes through three separate angles —30, 60, and 90 degrees —and repeat the CNN training. 
 
Fig.~\ref{fig:boxplot} presents boxplots that quantify the deviations between the CNN-predicted values and the actual magnetic field strength. These deviations are quantified by calculating the absolute differences in the magnetic field strength distribution, $|B^{\rm CNN} - B|$, denoted as $\sigma_B$. After introducing rotation angles, we observe no significant changes in $\sigma_B$. The uncertainty ranges from \SI{0.1}{\micro G}  to \SI{1}{\micro G} , with median values of $\sigma_B$ approximately \SI{0.22}{\micro G} , \SI{0.1}{\micro G} , and \SI{0.01}{\micro G}  for $\beta = 100$, 200, and 500 conditions, respectively. The CNN performs best for the $\beta = 500$ case, where the maximum uncertainty remains below \SI{0.5}{\micro G} . Overall, the CNN model’s efficiency in predicting magnetic field strength appears to be largely insensitive to variations in view angles.

\subsubsection{Noise Effect}
Noise is an unavoidable aspect of observational data that can potentially influence the predictions made by CNN models. To assess this effect numerically, we introduced Gaussian noise \footnote{It should be noted that other sources of noise may also happen in observation and need further study.} into the synchrotron intensity maps utilized for training our CNN. The amplitude of the noise was varied to represent 1\%, 20\%, 33.3\%, and 50\% of the mean intensity of the maps, corresponding to the Noise Ratio(NR) of 1/10, 1/5, 1/3, and 1/2, respectively. 

Fig.~\ref{fig:boxplot} presents boxplots that illustrate the deviations between the CNN-predicted magnetic strengths and the actual values under these different noise conditions. In a noise-free environment, the median values of $\sigma_B$ are approximately \SI{0.1}{\micro G}. With the introduction of noise, the uncertainties in predictions increase significantly. The median values of $\sigma_B$ rise to approximately \SI{1}{\micro G}  for an NR of 1/2, \SI{0.5}{\micro G}  for an NR of 1/5, and \SI{0.3}{\micro G}  for an NR of 1/10. The increase in uncertainties correlates consistently with the intensity of the noise, demonstrating that higher levels of noise lead to greater prediction errors. 
Notably, the maximum uncertainties also escalate with stronger noise levels, indicating that the model's predictive accuracy deteriorates under high noise conditions. 
We also conducted tests with NR$\ge1/2$; however, at these extremely high NR levels, the magnetic field information was almost completely obscured by noise, making it impossible to make reliable predictions.

\subsection{Prospects of the CNN-predicted magnetic field strength}
Understanding the magnetic field within galaxy clusters is crucial for addressing fundamental questions concerning cosmic ray acceleration mechanisms, the amplification of magnetic fields by turbulent dynamos, and the properties of dark matter candidates such as axion-like particles.

Building on the insights that the morphology of observed synchrotron structures encodes information about magnetic field strength \citep{2024arXiv240407806H,2024arXiv241009294H}, we propose utilizing Convolutional Neural Networks (CNNs) to extract these morphological features from synchrotron emissions and estimate magnetic field strengths in galaxy clusters. This approach is viable across different types of galaxy clusters, provided that suitable training datasets are available.

\subsubsection{Cosmic ray acceleration and transport}
An accurate characterization of magnetic field properties is essential for understanding the physics of cosmic rays (CRs). For example, mechanisms such as diffuse shock acceleration \citep{1978MNRAS.182..147B, 2000IAUS..195..291A, 2014IJMPD..2330007B, 2022ApJ...925...48X} and turbulent second-order Fermi acceleration \citep{2001MNRAS.320..365B, 2014IJMPD..2330007B} are widely recognized for their roles in CR acceleration within galaxy clusters. Both the strength and orientation of magnetic fields are critical factors influencing acceleration efficiency. Additionally, the diffusion coefficients for CRs, which are correlated with the medium's magnetization level, are influenced by the characteristics of turbulent magnetic fields \citep{2008ApJ...673..942Y, 2013ApJ...779..140X, 2022MNRAS.512.2111H}. The magnetic field strengths measured by CNNs can provide pivotal insights into the acceleration and transport processes of CRs

Moreover, the CNN approach does not presuppose equipartition between accelerated electrons and magnetic fields; it relies solely on the morphology of synchrotron emissions. Comparing the magnetic field strengths derived from equipartition and those predicted by CNNs offers a way to further constrain the relativistic electron number density and to explore the energy distribution among electrons, protons, and magnetic fields (see Fig.~\ref{fig.Beq}).

\subsubsection{Magnetic field amplification}
Magnetic field amplification during galactic mergers remains poorly understood. Previous numerical studies \citep{1999ApJ...518..594R, 2008ApJ...687..951T, 2018SSRv..214..122D, 2018MNRAS.474.1672V} have suggested that magnetic fields evolve in concert with cluster dynamics: the fields are stretched and stirred, amplified by large-scale bulk flows along the merger axis. A small-scale turbulent dynamo further amplifies the magnetic field at subsequent stages. These predictions have been explored by earlier measurements from Rotation Measure grids \citep{2004A&A...424..429M} and measurements of magnetic field orientation via synchrotron intensity gradients \citep{2024NatCo..15.1006H}. Our CNN methodology offers new perspectives for mapping the magnetic field strength within clusters, facilitating a detailed comparison between the numerical predictions of merging clusters and observational data.

\subsubsection{Cosmology: constrain axion-like particles}
Axion-like particles (ALPs) are a class of pseudoscalar particles that can generically couple to photons, making the oscillations between photons and ALPs in the presence of external magnetic fields possible \citep{2013ApJ...772...44W}. Efforts have been made to explore ALP parameters, particularly within the context of the Perseus cluster \citep{2020PhLB..80235252L}. The distinctive oscillatory features in astrophysical spectra caused by photon-ALP mixing can be characterized by magnetic field, providing a method to constrain ALP parameters through observations in several Galactic sources relevant for ALP masses ($\sim10^{-9}-10^{-7}$eV) \citep{liang2019constraints, majumdar2018search}. Knowledge of magnetic field orientation and strength is crucial for placing stringent constraints on ALP parameters.

\subsection{Comparison with earlier work}
The exploration of magnetic fields using CNNs is advancing rapidly. \cite{2024MNRAS.52711240H} initially introduced a CNN model designed to estimate the 3D structure of magnetic fields, including the orientation of the POS magnetic field, the inclination angle, and the magnetization level of the medium in molecular clouds. Subsequently, \cite{2024arXiv240407806H} extended this approach to synchrotron emission observations to trace the 3D magnetic field.

Previous studies primarily focused on estimating the medium's magnetization rather than the magnetic field strength itself. Building upon this foundation, our study expands the application of CNNs to the latter objective, utilizing synchrotron emission observations. Our CNN model is trained using MHD simulations of galaxy clusters, incorporating complex physical processes such as cluster mergers and sloshing motions. We demonstrate the CNN's capability to process these complicates and estimate magnetic field strengths.


\section{Conclusion}
\label{sec:con}
In this study, we present a method that utilizes Convolutional Neural Networks (CNNs) in conjunction with synchrotron emission observations to estimate the magnetic field strength in galaxy clusters. The CNN approach leverages the distinctive features of magnetic field variations within observed synchrotron structures, offering a novel technique for extracting these variations. Our key findings are summarized as follows:
\begin{enumerate} 
\item We designed and implemented a CNN model capable of extracting magnetic field strength from synchrotron emission maps in Perseus-like cool-core galaxy clusters.
\item By training the model on MHD simulations of galaxy cluster mergers, we found that the median uncertainties for the CNN-estimated magnetic field strength are approximately \SI{0.22}{\micro G}, \SI{0.1}{\micro G}, and \SI{0.01}{\micro G} for $\beta=100$, 200, and 500 conditions, respectively. 
\item The model's robustness against noise and varying viewing angles was assessed, demonstrating that, with sufficient training, it is largely insensitive to these factors, ensuring reliable performance under a wide range of observational conditions.
\item We compared the magnetic field strengths estimated by the CNN with those derived using the equipartition method. Our CNN approach, which does not rely on equipartition assumptions, provides new insights into the constraints on relativistic electron number density and cosmic ray acceleration mechanisms.
\end{enumerate}

%

\begin{acknowledgments}
Y.H. acknowledges the support for this work provided by NASA through the NASA Hubble Fellowship grant No.~HST-HF2-51557.001 awarded by the Space Telescope Science Institute, which is operated by the Association of Universities for Research in Astronomy, Incorporated, under NASA contract NAS5-26555. A.L. acknowledges the support of NSF grants AST 2307840 and ALMA SOSPADA-016. This work used SDSC Expanse CPU, NCSA Delta CPU, and NCSA Delta GPU through allocations PHY230032, PHY230033, PHY230091, PHY230105, and PHY240183 from the Advanced Cyberinfrastructure Coordination Ecosystem: Services \& Support (ACCESS) program, which is supported by National Science Foundation grants \#2138259, \#2138286, \#2138307, \#2137603, and \#2138296. 
\end{acknowledgments}

\vspace{5mm}
\software{Python3 \citep{10.5555/1593511}; TensorFlow \citep{tensorflow2015-whitepaper}}


\newpage
\appendix
\section{CNN training and test using MHD turbulence simulations}
\label{appendix}

We test the CNN model's performance in estimating the magnetic field strength using MHD turbulence simulations. The MHD turbulence simulations were generated from the AthenaK code \citep{2024MNRAS.527.3945H,2024arXiv240916053S}. We solve the ideal MHD equations within an Eulerian framework, complemented by isothermal and periodic boundary conditions. Kinetic energy is solenoidally injected at wavenumber 2 to get a Kolmogorov-like power spectrum. The computational domain was discretized into a $792^3$ cell grid, with numerical dissipation of turbulence occurring at scales between approximately 10 to 20 cells. Initial conditions for the simulations featured a uniform density field and a uniform magnetic field. Characterization of the scale-free turbulence within the simulations was achieved through $M_s$ and $M_A$. To explore the physical conditions of galaxy clusters, we use $M_s \approx 1$ and $M_A\approx1, 2, 3, 4$, and 5, corresponding to $\beta=2, 8,18,32,$ and 50, respectively. 

\begin{figure*}
    \centering
    \includegraphics[width=1.0\linewidth]{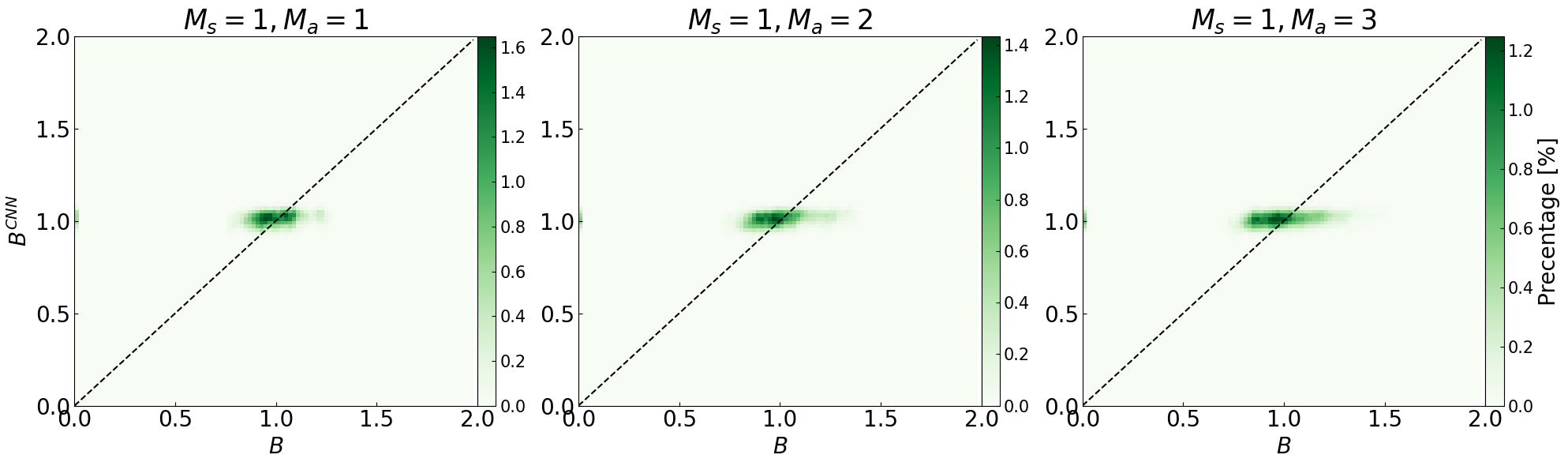}
    \caption{
2D histograms of the CNN-predicted normalized magnetic field strength $B^{\rm CNN}$ and the corresponding actual normalized density-weighted $B$. The magnetic field strength is normalized by the global mean value. The three panels represent three physical conditions of $M_s=1, M_A=1$ (left), $M_s=1, M_A=2$ (middle), and $M_s=1, M_A=3$ (right). The dashed diagonal line references the ideal scenario, where the predicted values match actual values.}
\label{predictions_pure}
\end{figure*}

Fig.~\ref{predictions_pure} presents 2D histograms comparing CNN predictions with actual magnetic field strengths derived from simulations with $M_A \approx 1, 2,$ and 3. The CNN predictions show a strong statistical correlation with the true magnetic field strengths, especially when normalized magnetic strength is considered instead of the absolute value. This indicates the potential of estimating magnetic field strength based on synchrotron emission maps.

\newpage
\bibliography{sample631}{}
\bibliographystyle{aasjournal}



\end{document}